\documentclass[conference]{IEEEtran}
\IEEEoverridecommandlockouts
\usepackage{times}
\usepackage{soul}
\usepackage{url}
\usepackage[hidelinks]{hyperref}
\usepackage[utf8]{inputenc}
\usepackage[small]{caption}
\usepackage{graphicx}
\usepackage{subcaption}
\usepackage{amsmath}
\usepackage{booktabs}
\usepackage{algorithm}
\usepackage{algorithmic}
\usepackage{xcolor}
\usepackage{amssymb}
\usepackage{cite}
\usepackage{amsmath,amssymb,amsfonts}
\usepackage{algorithmic}
\usepackage{graphicx}
\usepackage{textcomp}
\def\BibTeX{{\rm B\kern-.05em{\sc i\kern-.025em b}\kern-.08em
    T\kern-.1667em\lower.7ex\hbox{E}\kern-.125emX}}
\graphicspath{{Figs/}}
\begin{document}

\title{News Labeling as Early as Possible: Real or Fake?\\
}
\author{
\begin{tabular}[t]{c@{\extracolsep{8em}}c} 
Maryam Ramezani\textsuperscript{*}  & Mina Rafiei\textsuperscript{*} \\
\thanks{\textsuperscript{*}These authors contributed equally to this work.}
Department of Computer Eng & Department of Computer Eng \\ 
Sharif University of Technology & Sharif University of Technology \\
Email: maryam.ramezani@sharif.edu & Email: mrafiei@ce.sharif.edu
\end{tabular}\\\\
\begin{tabular}[t]{c@{\extracolsep{8em}}c} 
Soroush Omranpour  & Hamid R. Rabiee \\
Department of Computer Eng & Department of Computer Eng \\ 
Sharif University of Technology & Sharif University of Technology \\
Email: omranpour@ce.sharif.edu & Email: rabiee@sharif.edu
\end{tabular}
}


\maketitle
\begin{abstract}
Making disguise between real and fake news propagation through online social networks is an important issue in many applications. The time gap between the news release time and detection of its label is a significant step towards broadcasting the real information and avoiding the fake. Therefore, one of the challenging tasks in this area is to identify fake and real news in early stages of propagation. However, there is a trade-off between minimizing the time gap and maximizing accuracy. Despite recent efforts in detection of fake news, there has been no significant work that explicitly incorporates early detection in its model. In this paper, we focus on accurate early labeling of news, and propose a model by considering earliness both in modeling and prediction. The proposed method utilizes recurrent neural networks with a novel loss function, and a new stopping rule. Given the context of news, we first embed it with a class-specific text representation. Then, we utilize the available public profile of users, and speed of news diffusion, for early labeling of the news. Experiments on real datasets demonstrate the effectiveness of our model both in terms of early labelling and accuracy, compared to the state of the art baseline and models.
\end{abstract}


\section{Introduction}
With increased usage of online social networks, social interaction of users are increasingly growing. Users of social networks share ideas, thoughts, feelings, knowledge, news and advertisement via actions such as posting, commenting, liking and reposting. Through these networks, people are exposed to large amounts of information and can broadcast them on different channels on a daily basis. Verifying the validity of news items is an important issue in information diffusion over social networks. It only takes a short amount of time for a piece of news related to market products, social opinions and political issues to propagate among people across the world. The speed of propagation in cyberspace allows users to spread deceptive and valid news in different communities. Fake news can have destructive effects on human beliefs, decisions and behaviors. Deceitful users may intentionally
publish fake news to gain profit for themselves. This phenomenon has negative consequences, such as economic and social turbulences, false advertising, or the orientation of the general mind towards a particular opinion. Therefore, one of the challenging tasks in this area is to classify fake and real news in early stages, and to inform users about the validity of news.

{In recent years, various works have been done in detecting the truth of information using different terms such as rumor, hoaxes, misinformation, disinformation, claims, and fake news \cite{alrubaian2019credibility, Shu2019mitigation}. Rumor is widespread pieces of information which are unverified at an early stage and then can remain unverified for a long time or turn out to be a false or true statement. Rumor can be a believed and unintended false information that may be corrected during propagation (misinformation) or intentionally false information with the goal of misleading people (disinformation). Rumor detection is classifying that an event is a rumor or not. By definition, fake news is a type of hoax spreading in traditional news channels or via online social networks. In other words, fake news is disinformation and always false during its propagation life. Although in the early works the terms mentioned above have been used instead of each, here we use term “news” because we are classifying the verified rumors which are deliberately false and misleading (fake) or real \cite{Zubiaga:2018:DRR:3186333.3161603}.}

Distinguishing between real and fake news has always been a challenge, however it is not feasible to automatically detect them in traditional diffusion channels. Today, there are some fact checking sites and plug-ins for debunking such as \href{FactCheck.org}{FactCheck.org} which monitor political claims, and \href{Snopes.com}{Snopes.com} focusing on news stories. However, manual or late detection of fake news are drawbacks of these sites. Automatic detection has been largely studied in the past by using various learning approaches. 
Different works provide reviews for the fake news methods and classify them into different categories. The authors in ~\cite{Shu:2017:FND:3137597.3137600,DBLP:journals/corr/abs-1807-03505} categorize the various type of data for detection of fake news including text content, visual contents, user based and social context features. The authors in \cite{DBLP:journals/corr/abs-1807-03505} divide the proposed methods into handcrafted features, propagation-based and neural networks approaches. The authors in \cite{DBLP:journals/corr/abs-1812-00315} list the fundamental theories to study fake news and study various perspectives of fake news research. One of the main future directions in fake news research is early fake news detection \cite{Shu:2017:FND:3137597.3137600,DBLP:journals/corr/abs-1812-00315}.  Despite various fake news detection methods, there has been no significant work focusing on earliness, and all of prior works have confined earliness just in evaluating the performance of their algorithms. 
\begin{figure*}[]
\centering
\includegraphics[width=0.8\textwidth]{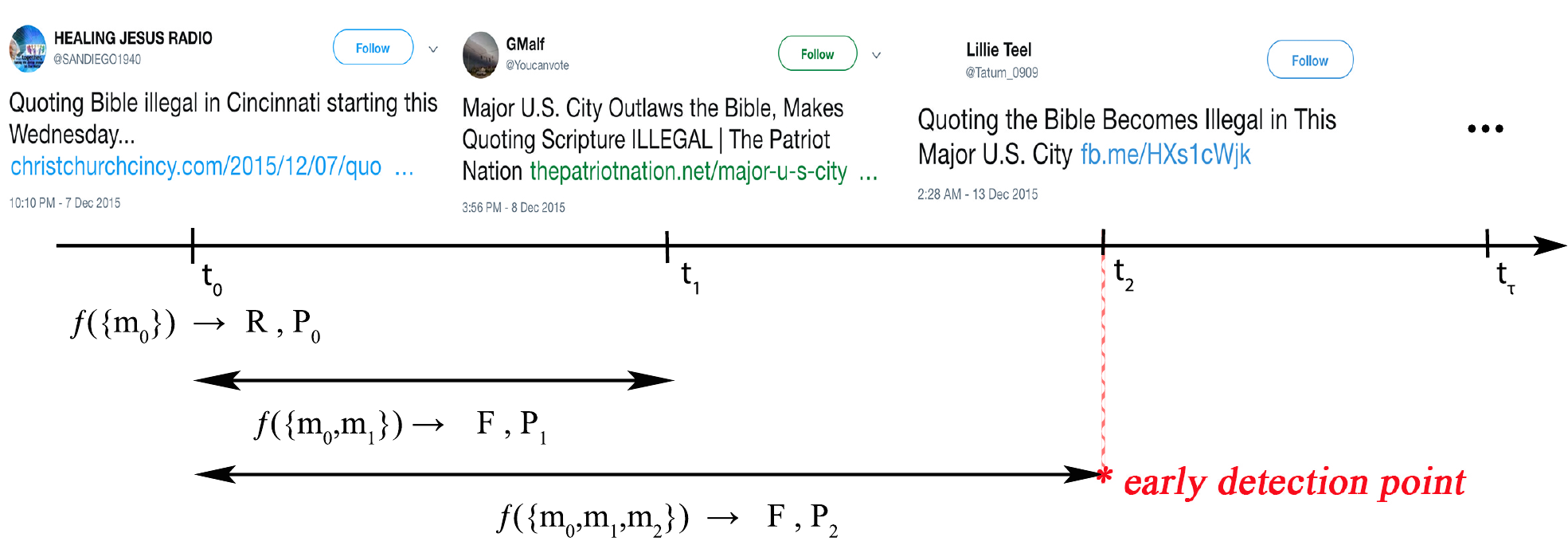}
\begin{center}
\caption{\label{fig:IntroExam} Early labeling of news stream. A label $F$ is assigned to the sequence at early time stamp $t_2$ which has high classification probability $P_2$ same as the label when using the total $t_\tau$ steps of stream.}
\end{center}
\end{figure*}

In this paper, we focus on the early labeling of news. We represent a news dissemination by a time series sequence. At each time step, we estimate the probability of labeling the stream. Clearly, with more incoming data, more accurate detection can be achieved. Our aim is to label the stream with high probability of being fake or real, as early as possible (Figure \ref{fig:IntroExam}). This early detection can help to prevent the fakes before extensive expansion through the network. We propose a novel loss function for RNN based learning to train the model by considering early classification and introduce a new stopping rule for the testing phase. In summary, we make the following contributions:
\begin{itemize}
\item We propose a new loss function for early classification of the news streams. 
\item We utilize our loss function in training a RNN model for spread of news along time and do not assume any prior information or distribution.
\item We introduce a rule for making decisions about labeling the news streams or waiting for more data.
\item Our proposed method utilizes the public and available information of news that can be accessed with no effort.
\item We achieve greater accuracy as compared with the start of the art methods in detecting the fake news in an early stage.  
\end{itemize}

The rest of this paper is organized as follows. Section 2, provides the related works. Problem formulation and the proposed method is discussed in Section 3. Section 4 demonstrate the analytic performance of the proposed method, and its effectiveness compared to the baselines with experimental results on real datasets. Finally, we conclude the paper and discuss the future works in section 5.

\section{Related Works}
Works related to news identification can be categorized based on different aspects. It is necessary to mention that not all of these researches are about fake news, and may include related topics such as rumours and misinformation.
\begin{enumerate}
\item{Approach of classification}: They utilize machine learning algorithms with hand-crafted features \cite{zhao2015enquiring,wu2015false,ma2015detect,Sampson:2016:LIS:2983323.2983697,wu2017gleaning} or a deep neural network \cite{ma2016detecting,wu2018tracing,csi2017,aaai2018,zhang2018fake}. 
\item {Input Features}: Approaches in the literature use different features of input data including content and network. Features can be one of the four types: 

\textbf{Linguistic} cues such as word level features, emotion and semantic symbols, sweer words and special characters usually used in natural language processing can help to distinguish the fake statements \cite{castillo2011information,qazvinian2011rumor,gupta2014tweetcred,mitra2017parsimonious}. However, as the design of fake news become more intelligent, fake linguistic features are selected more carefully to deceive detection models.

\textbf{Linguistic and Textual Sequence} is generated by different comments of users on the original news over the time. A question mark presence on repost of a news indicates the doubt of users about the validity of information. \cite{zhao2015enquiring} use these question answering and enquiry phrases as early signal of rumour detection with SVM and decision tree, while these user actions are inadequate at early stage and require extracting the best signals as features. As a first attempt of using deep networks in this field, \cite{ma2016detecting} propose three methods based on RNN, single layer LSTM and GRU, and multi-layer GRU using linguistic sequences. To focus on important relevance, \cite{chen2017call} add a soft attention mechanism to the previous model.

\textbf{User Sequence} is a series of profile and information about the users participating in a news diffusion. \cite{wu2018tracing} embed the users in sequence of news spreading with graph based algorithm. Then train a LSTM-RNN model to classify the labels. Access to the network structure is the heavy information required by this model. In an early research, \cite{aaai2018} first encode the user sequence by CNN and GRU, independently. Then, output of these networks are concatenated and fed to a feed forward network for classifying. This paper emphasizes that within five minutes after the start point of a propagation, it can detect fake news around $90\%$ of accuracy. This conclusion is a bit unclear, because it need to average the length of diffusion for first five minutes in each dataset and then run the model with the calculated length of sequences. Whereas, dissemination processes are vary in length and this model can not optimize the earliness for each sequence and early detection is not included automatically in model training. 

\textbf{Propagation Structure}
Diffusion graphs of real and fake news are structurally different. These trees can be utilized in a SVM classifier as simple kernel \cite{P17-1066} or a hybrid kernel composed of radial basis and a random walk kernel \cite{wu2015false} .

Using multiple different features is a good way to achieve high classification accuracy. Combination of linguistic and user sequence has been taken into account in \cite{ma2015detect}. \cite{csi2017} is another hybrid model in which Context (linguistic) and response of users (linguistic sequence) are given to a RNN. Simultaneously, a fully connected layer measures the score of user sequence. Finally, the sequence is labled by integration of these two outputs. On the other hand, a group of previous works employ additional inference information by constructing a tripartite graph between users, news, topics or publishers \cite{zhang2018fake,shu2017exploiting}.
\item{Detection speed}: Indeed, what is the detection algorithm or which input features it used, if a model utilize all the information from the beginning to the end of the diffusion is a late model, and if it detect the label from partially primary information we call it an early model. 
\end{enumerate}
Even though in many of related works an early concern is mentioned, but the earliness is not included in any of previous models and just is used as a performance metric for evaluation with checking the accuracy against percentage of training data \cite{wu2017gleaning} or by varying the maximum number of observed posts in a test sequence \cite{P17-1066,Sampson:2016:LIS:2983323.2983697,shu2017exploiting,ma2016detecting,aaai2018,Guo:2018:RDH:3269206.3271709,Jin:2016:NVE:3016100.3016318,Yu:2017:CAM:3172077.3172434}.

\section{News Early Classification}
\begin{figure}[t]
\centering
\includegraphics[width=0.48\textwidth]{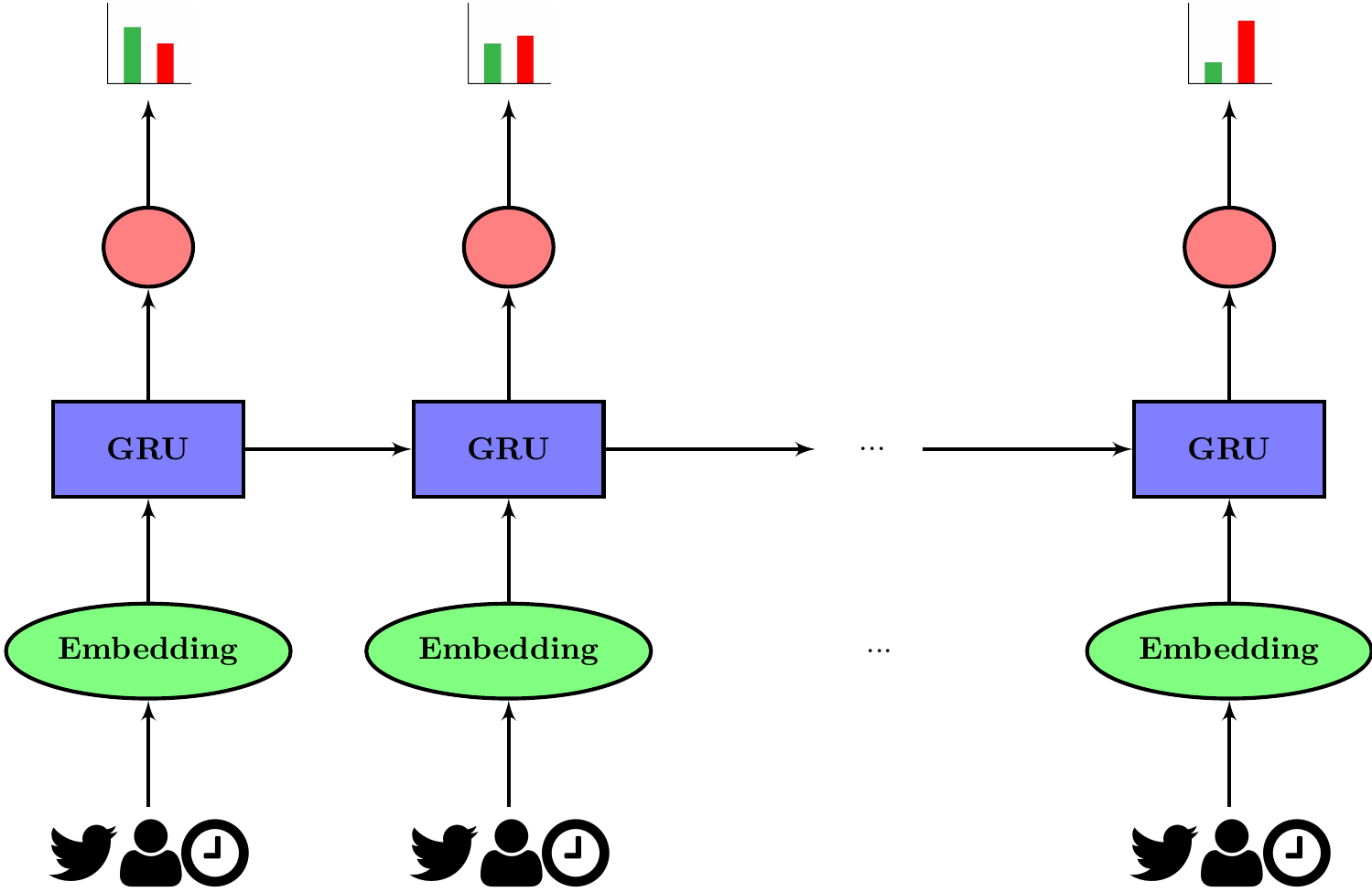}
\begin{center}
\caption{\label{fig:arch} Model Architecture of NEC.}
\end{center}
\end{figure}
In this section, first we state the problem formulation. Then, we describe our News Early Classification (NEC) method in details. NEC can label the stream of a news to fake or real class as early as possible, by utilizing the recurrent neural network.

\subsection{Problem Formulation}
A \textit{news event} is a widespread news that can be tracked by specific topic and keywords on social networks. Let $E = \{e_1, e_2, e_3, ..., e_{N} \}$ be the set of different news events
disseminated daily among users. Each news event $e_i$ is provided by the set of messages $m_i$s that social users post and share over the time. In the same way, $e_i(t)=\{m_{i1},m_{i2},...,m_{it}\}$ is the streaming set of a news after $t$ steps of spreading. Let $\tau$ be the maximum length of the streaming of a news, then $e_i$ is defined as $e_i(\tau)$. Then, the news labeling problem on each $e_i$ is learning a binary classification $f(e_i)=\ell$ such that $\ell \in \{0,1\}$ where $0$ means real label and $1$ means fake.

Here, we aim to learn an early classifier function $f_{\Theta}$ that minimizes a supervised loss. In the real time classification step for a message stream $e_i(t)$, the learned model will assign a label $\ell$ with probability $P$ till the time step $t$,
or it waits for more incoming messages. To this end, we define the following function and learn the parameters $\Theta$. Let $S=\{D,W\}$ be the corresponding \textit{Detection} and \textit{Waiting} states, then:
\begin{equation}\label{eq:f_teta}
{f_{\Theta}(e_{i}(t))}  =(S,{\ell},P)
\end{equation}
It is clear that when $S=W$, the ${\ell}$ and $P$ are null. 
\subsection{Model Description}
Time is one of the fundamental elements in news diffusion. In streaming a news, the time and ordering of related messages are two important keys, and we have a stream time series $X=\{X_1,X_2,...,X_t,...\}$ sorted in ascending order of time. Based on the sequential dependency in diffusion process, and our goal to model the probability of output label at any time step $X_t$, we employ the Recurrent Neural Networks (RNNs) to learn the interaction between news label and characteristics of $X$ at early stage. The overall architecture of NEC is illustrated in Figure \ref{fig:arch}. Formally, $X_t$ is given to GRU (Gated Recurrent Unit) \cite{W14-4012}, an improved version of RNN, to obtain the hidden state $h_t$ as output:
\begin{equation}\label{eq:ht}
h_t=(1-z_t).h_{t-1}+z_t.\tilde{h}_t
\end{equation}
with:
\begin{equation}
 \begin{array}{l}
 {\tilde{h}_t}=tanh(W_hX_t+U_h(r_t.h_{t-1}))\\
 z_t=\sigma(W_zX_t+U_zh_{t-1})\\
 r_t=\sigma(W_rX_t+U_rh_{t-1})
\end{array}
\end{equation}
Then $h_t$ will fed into a linear layer for reduction to two dimensions ($ \hat{y_t}$), and generating the probability of binary classification by using a soft-max activation function.
\begin{equation}\label{eq:pt}
 \begin{array}{l}
 \hat{y_t}=Wh_t+b\\
\\
P_t=\dfrac {e^{ \hat{y_t}}}{{e^{{ \hat{y_t}}{[0]}}}+{e^{{ \hat{y_t}}{[1]}}}}
\end{array}
\end{equation}

Tracing the diffusion path and inferring the network structure are two expensive features that cannot be obtained in the early stage of news propagation, while using textual and user sequence features act consistently good even in the beginning of diffusion, when the information is partially available \cite{kwon2017rumor}. Here, $X_t$ contains the information of a message $m_{it}$ in stream time series of event $e_i$, represented by $X_t=(C_t,U_t,DT_t)$, where $C$, $U$ and $DT$ are input features representing message context, user information and message post time, respectively. \\
\textbf{Message context ($C$)}: The words distribution in two fake and real classes are different. A word can have two different meanings using in a real news or displaying in a fake article. As \cite{zubiaga2018learning} states, class-specific word training can improve the text embedding in hoaxes detection. NEC utilized a Doc2Vec embedding \cite{doc2vec2014} on fake and real news to construct feature vectors $C^{F}_{t}$ and $C^{R}_{t}$ for textual analysis, independently. Therefore, $C_t$ is:
\begin{equation}\label{eq:Ct}
C_t=concatenate(C^{R}_{t},C^{F}_{t})
\end{equation}\\
\textbf{User information ($U$)}: User social profile is the simplest and most accessible information that we can have about the users in a social network. If a message is public and displayed, the user profile can also be obtained. Followers count as the popularity and influence of user, account verification for invoking a trust, date that user joined the social network, Geo status, length of user name and user description, following and status count are eight available features of a user profile as utilized in \cite{aaai2018}.\\
\textbf{Message post time ($DT$)}: False news diffused significantly faster than real news \cite{Vosoughi1146}. At time step $t$, if the post time of message is $T_t$, $DT_t=T_t-T_0$ can be considered as a sign of news diffusion speed.
\begin{algorithm}[tb]
\caption{News Early Classification Algorithm}
\label{alg:algorithm}
\textbf{Input}: news message stream $x=<m_1,m_2,..>$\\
\textbf{Output}: label of news
\begin{algorithmic}[1] 
\STATE  seq = [] 
\STATE  t = 0 
\WHILE{ t $\leq $ threshold }
\STATE seq.add( x[t] )
\STATE state , label = $f_\Theta$ (seq)
\IF {state == D }
\STATE break  
\ENDIF
\ENDWHILE
\STATE \textbf{return} label
\end{algorithmic}
\label{alg:stop}
\end{algorithm}

\subsection{Training step}
We are given a set of $N$ news events $\{e_1, e_2, e_3, ..., e_{N}\}$ with their corresponding labels $\{y_1,y_2,y_3,...,y_{N}\}$. 
Since the final life time ${\tau}$ for a news $e_i$ is available in training data, we have the time series stream $X^{(i)}=\{X^{(i)}_{1},X^{(i)}_{2},...,X^{(i)}_{\tau}\}$. For training the $\Theta$ parameters of Equation \ref{eq:f_teta}, RNN tries to minimize a loss function. Cross entropy is a good loss for binary classification. For a focussed learning on earliness, we modify the cross entropy by weighting the loss at each time step $1\preceq t \preceq \tau$ with $\gamma(t)$ as follows:

\begin{equation}\label{eq:loss}
\begin{array}{l}
\mathbb{L}({\Theta}|X,Y)
=\dfrac{1}{N}\sum\limits_{i = 1}^N \sum\limits_{t = 1}^\tau

-\gamma(t)
[
{y_{t}}^{(i)} log P({\hat{y}}^{(i)} |{X_{1:t}}^{(i)},\Theta)\\
+(1-{y}^{(i)})log(1-P({\hat{y_t}}^{(i)}|{{X_{1:t}}^{(i)}},\Theta))
]
\end{array}
\end{equation}

where $y_t$ is the ground truth label of the stream sequence of a news and $P({\hat{y_t}}|{X_{1:t}},\Theta)$ is the probability of our model detected label for time step $t$ given the learnable parameters $\Theta$, and received information from the start of propagation $X_{1:t}$.

The coefficient $\gamma(t)$ is used as a trade off between two inconsistent objectives including accuracy and earliness. Very high accuracy is not expected at the early steps of diffusion, because of limited amount of information available from the first stage. However, at later points in the time series sequence, classification should be more accurate. Therefore, we have a penalty term $\gamma(t)$ based on the observed length of sequence at each time step $t$ towards total life steps $\tau$ as shown in Equation \ref{eq:gamma}. By increasing this penalty term, one can force the model to label the news more accurately. 

\begin{equation}\label{eq:gamma}
\gamma(t)=-log(\dfrac{t}{\tau})
\end{equation}

\subsection{Classification step}
The early accurate detection point in different news streams may vary. Therefore, we utilize a stopping rule to evaluate if an accurate detection point is reached. To this end, at time step $t$ with probability of predicted label $\hat{y}$, the expression is defined as:
\begin{equation}\label{eq:stop}
[P(\hat{y}_t)\succ P(\hat{y}_{t-1})] \wedge [P(\hat{y}_t) \succ \alpha] \wedge [P(\hat{y}_t)-P(\hat{y}_{t-1}) \prec \gamma(t)]
\end{equation}
The first clause indicates that whether the sequence is at consistent point where the detected label probability will not decrease by observing more data during next time steps. The second clause defines a minimum value $\alpha$ on probability for labeling the news. The third clause prevents the model from requesting more information in order to increase the classification accuracy when observed time series is becoming undesirably long which can negatively effect the earliness. Algorithm \ref{alg:stop} shows the real-time procedure of labeling a news message stream that is spreading through the network. The algorithm starts as the first message is posted. At each time step $t$, the messages sequence till $t$ is fed to model and the output is checked in the stopping rule expression (Equation \ref{eq:stop}). If the expression evaluates as true, the state is set to D and loop of detection terminates with an output label. On the other hand, if the expression evaluates to false, the state is set to W, and loop will be continued for new incoming sequence containing the next message.

\section{Experimental Evaluation}
\subsection{Dataset Description}
We utilized two real-world datasets \cite{ma2016detecting}. The statistics of datasets are shown in Table \ref{tab:datasets}. Each dataset contains a set of news events labeled as fake or real.  A news event is consisted of a sequence of messages. Due to the terms of use of Twitter data, tweets and user information of this dataset are not published publicly. Therefore, we crawled the tweets and user data by using the Twitter API\footnote{https://dev.twitter.com/rest/public}.

\begin{table}[h]
	\centering
	\caption{ Datasets statistics }
	\label{tab:datasets}
	\begin{tabular}{|c|c|c|}  

		\hline
		Dataset & Sina Weibo & Twitter \\
		\hline
		\hline
		Real Events & 2351 & 493 \\
		Fake Events & 2313 & 498 \\
		Users & 2,746,818 & 381,541 \\
		 Average Number of Tweets per Event & 804 & 483 \\
		 Average news event Lifetime (hours) & 1,808 & 1,250 \\
		
		
		\hline
	\end{tabular}
	
\end{table}

\begin{table*}
	\centering
	\caption{ NEC results compared to competitive models on Sina Weibo and Twitter Dataset }
		\label{tab:results}
	\begin{tabular}{|c|ccc|ccc|ccc|ccc|}  
		\hline
		\multicolumn{1}{|c|}{}  &
		\multicolumn{6}{|c|}{Sina Weibo} &
		\multicolumn{6}{|c|}{Twitter}  \\
		\hline
			\multicolumn{1}{|c|}{}  &
			\multicolumn{3}{|c|}{Fake} &
			\multicolumn{3}{|c|}{Real} &
			\multicolumn{3}{|c|}{Fake} &
			\multicolumn{3}{|c|}{Real}  \\
		\hline

		model  & R & P & F & R & P & F&  R & P & F & R & P & F \\

		\hline
		NEC & \textbf{0.93} &	\textbf{0.87}&	\textbf{0.90}	 & \textbf{0.86}	& 0.92 & \textbf{0.89}
		& 0.93 & 0.64 & \textbf{0.76} & 0.53 & \textbf{0.90} & 0.67 \\
	
		\hline
		PPC &  1	& 0.78 &	0.88 &	0.77 &	\textbf{1} &	0.87 & 0.71 & 	\textbf{0.70} &	0.71 &	\textbf{0.68} &	0.69 & 	\textbf{0.69}\\
		
		\hline
		GRU2 &	0.73 &	0.67 &	0.70 &	0.65 &	0.71 & 0.68
		& 0.55 & 0.59 & 0.57 &	0.60 &	0.57 & 0.59  \\
		
		\hline
		CSI &  0.92 &0.73 &	0.81 &	0.67 &	0.90 &	0.77& \textbf{0.98} &	0.53&	0.68&	0.11&	0.86&	0.20 \\

		\hline
	\end{tabular}

\end{table*}

\subsection{Evaluation Metrics}
We evaluated the model performance from two points of view. First, the correctness of event labeling is evaluated with the metrics such as accuracy, precision, recall and F-measure. We report each metric per fake and real class because the goal is news labeling, rather than merely detecting fake news. Second, the model ability to label the events, as early as possible, is evaluated through the earliness metric. Earliness is measured by length ratio at which the model stops labeling. Therefore, the earliness is defined as:

\begin{equation}
	\displaystyle
	Earliness = \frac{1}{N} \sum_{i=1}^{N} \frac{t_i}{\tau_i}
\end{equation}
where $t_i$ denotes the stopping time step, and $\tau_i$ is the event maximum length.

\begin{figure}[!]
	\begin{subfigure}{0.45\textwidth}
		\centering\includegraphics[width=0.9\textwidth]{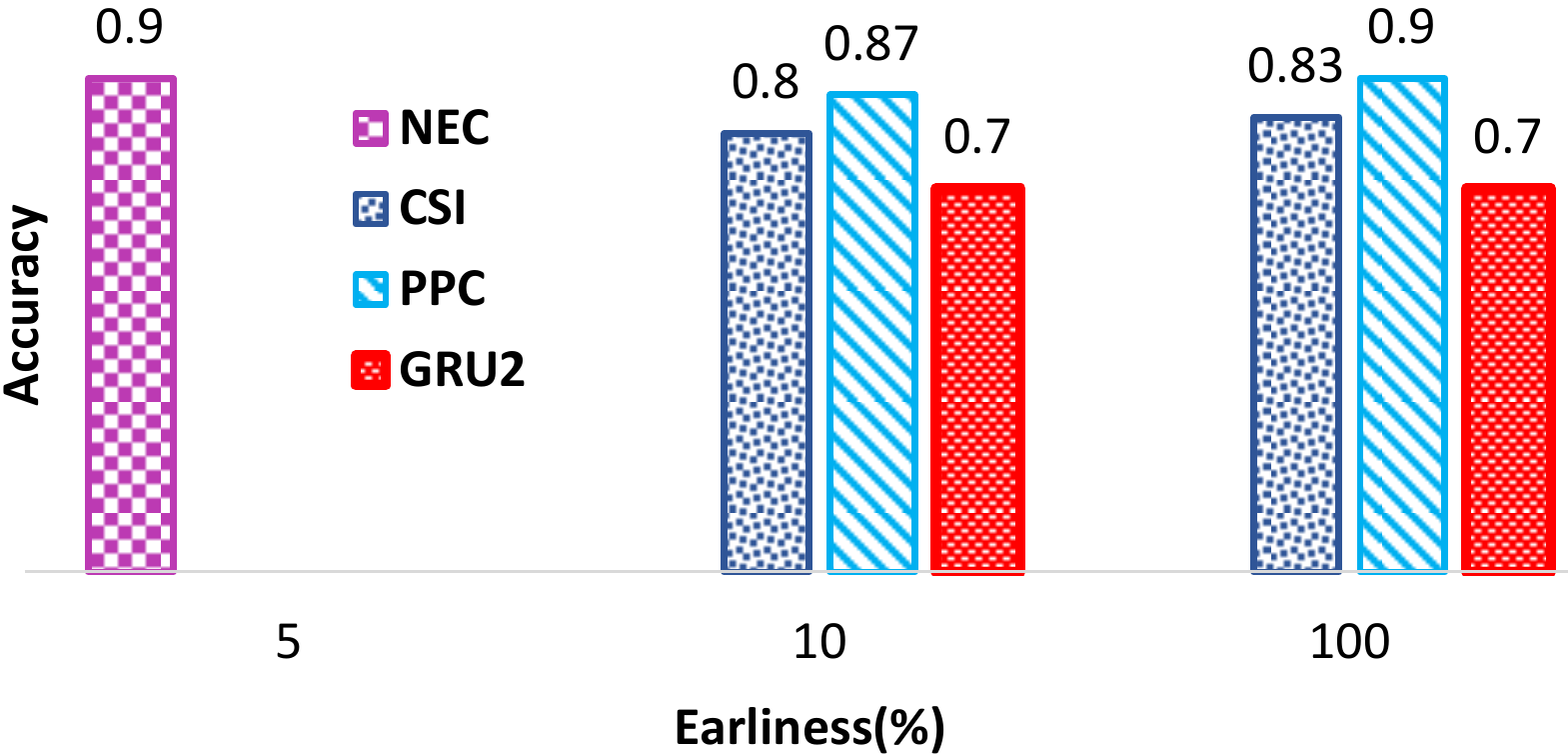}
		\caption{ Sina Weibo Dataset}
		\label{fig:acc_w}
	\end{subfigure}
	\begin{subfigure}{0.45\textwidth}
		\centering\includegraphics[width=0.9\textwidth]{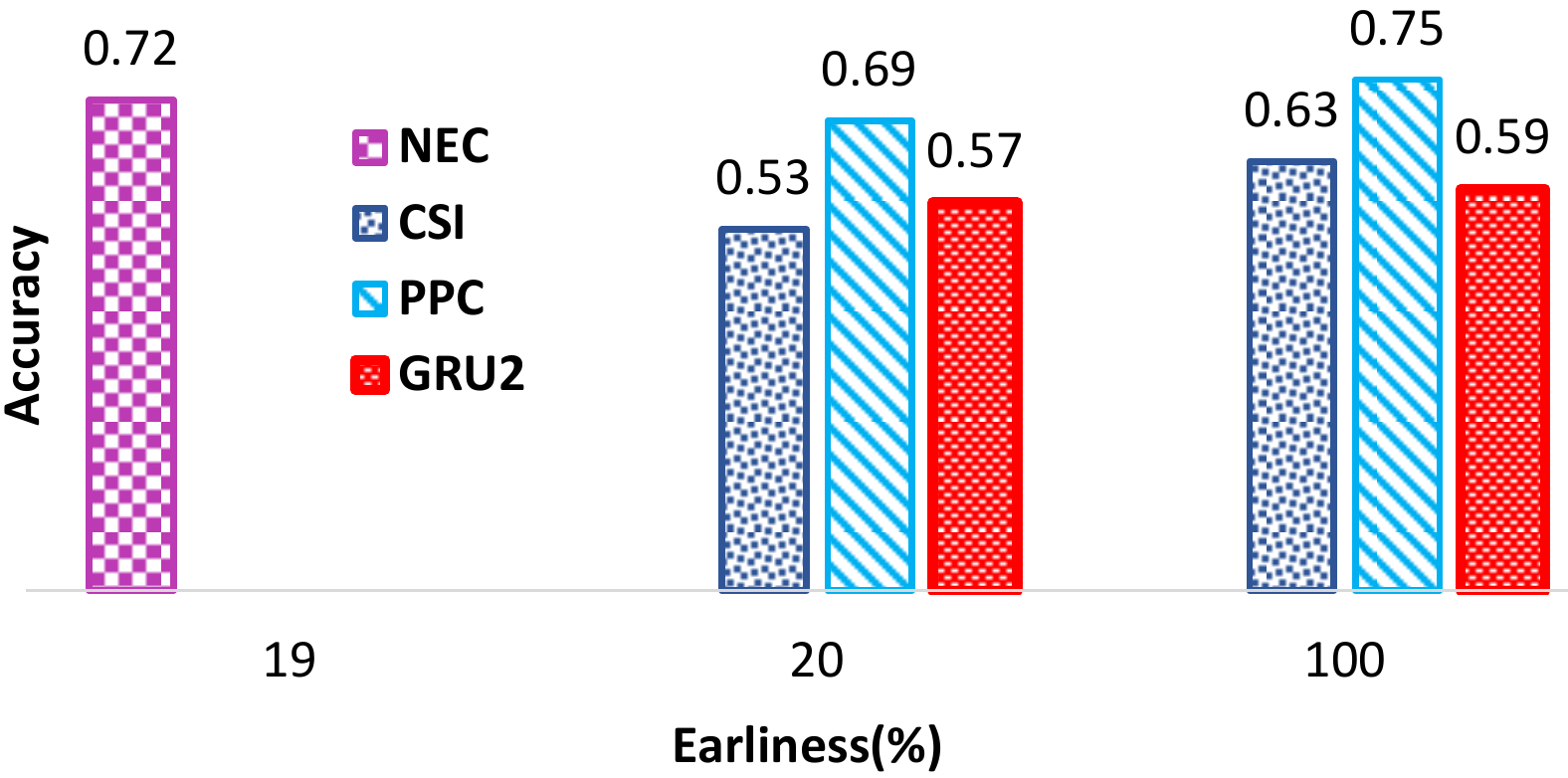}
		\caption{ Twitter Dataset}
		\label{fig:acc_t}
	\end{subfigure}
	\caption{Accuracy comparison in earliness point and 100\% of sequence}
	\label{fig:acc}
\end{figure}

\subsection{Experimental Setup}
We compared our model with the following baseline and models which were described in Section 2.
\begin{itemize}
\item \textbf{GRU-2} \cite{ma2016detecting}: As a baseline model.
\item \textbf{CSI} \cite{csi2017}: As a state of the art model in accuracy by utilizing both the user and text sequences. We used the implementation published by the authors\footnote{https://github.com/sungyongs/CSI-Code}. However, their released code used both train and test data for learning the embedding of text and user. For fair comparison, we have modified the proper parts of code to carry out the learning only with the train \footnote{https://github.com/s-omranpour/CSI-Code}.
\item \textbf{PPC} \cite{aaai2018}: As a state of the art model using both user and text sequence which claims to consider earliness. However, as discussed in Section 2, their earliness claim is not feasible. 
{The main differences between NEC and  PPC model are: 1) NEC utilize a combination of three categories of features containing user data (profile), text and time, but the PPC only uses the user’s data. 2) Our classification is real-time and gets the stream of messages as input and outputs the early detection point with labeling the sequence on that point (which can be different for different sequences), but the PPC model gets a sequence of fixed length messages (or pad it in case of being smaller) and outputs the label. 3) The architecture of the two models are also different. PPC uses CNN and RNN cell units but NEC only use RNN.}

\end{itemize}

\begin{figure}[]
	\begin{subfigure}{0.5\textwidth}
		\centering\includegraphics[width=0.7\textwidth]{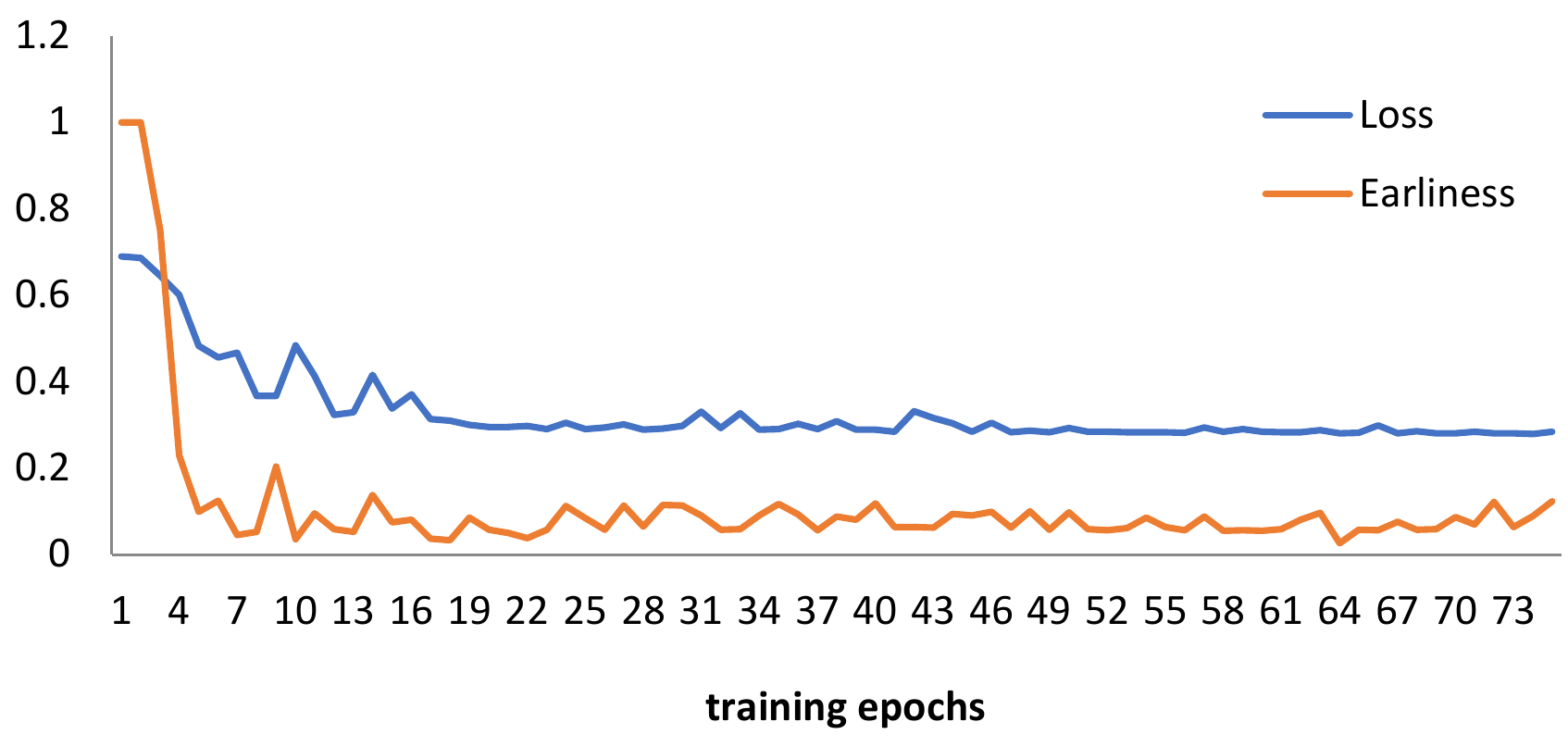}
		\caption{Sina Weibo Dataset}
		\label{fig:loss_w}
	\end{subfigure}
	\begin{subfigure}{0.5\textwidth}
		\centering\includegraphics[width=0.7\textwidth]{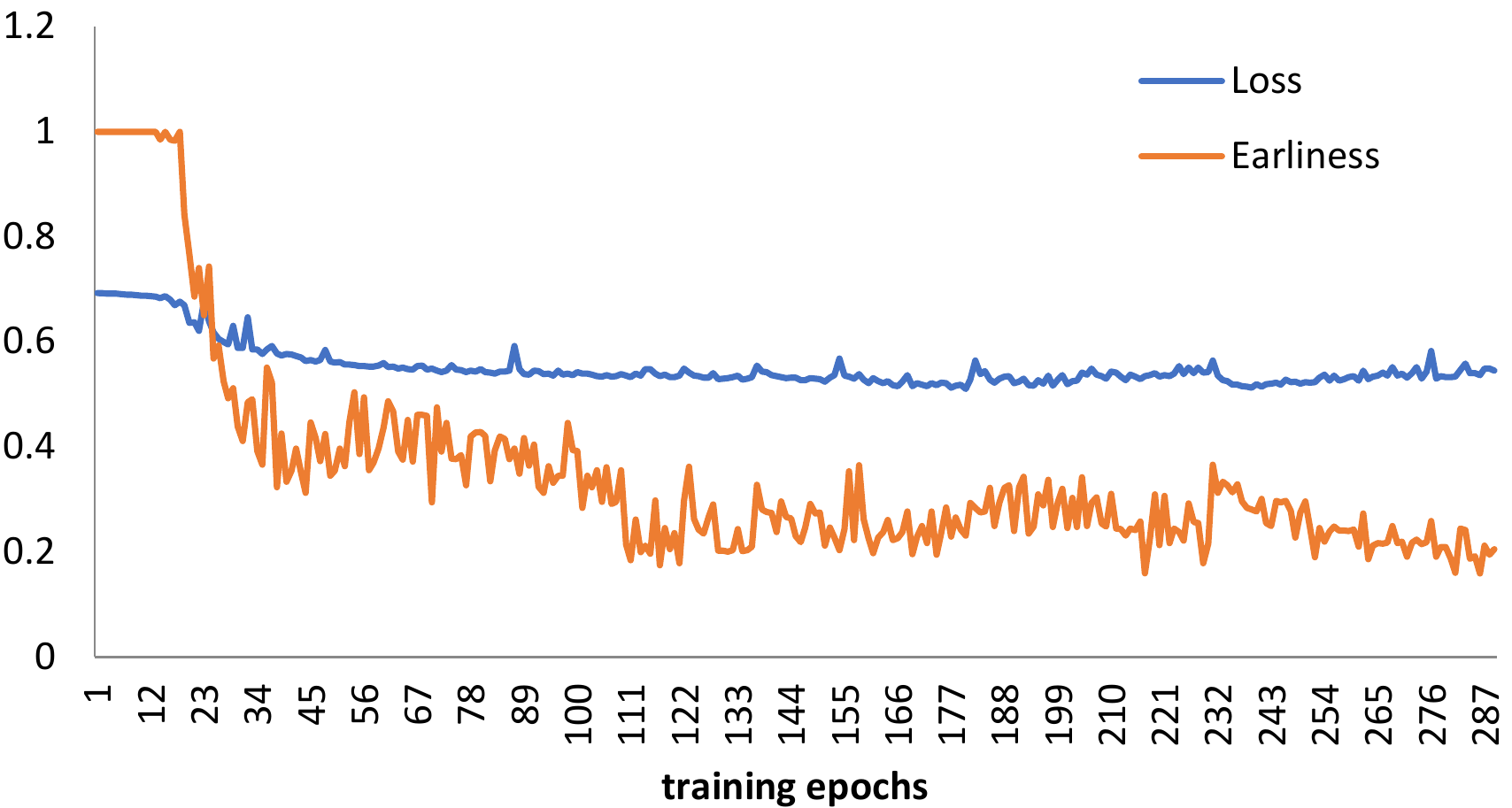}
		\caption{Twitter Dataset}
		\label{fig:loss_t}
	\end{subfigure}
	\caption{Loss and Earliness in Training epochs}
	\label{fig:lossepoch}
\end{figure}
We randomly selected $10\%$ of the events of each dataset as test sets, and the rest were divided with a ratio of 8 to 1 for training and validation sets, respectively. Models hyper parameters were selected based on how model performs on validation sets. We set hidden size of GRU to 32, the embedding vector size of each class of Doc2Vec to 50, and $\alpha$ to 0.7 and the maximum detecting time-step ($\tau$) to 100. The model was trained using gradient descent with the Adam optimizer \cite{adam}.
The learning rate was set to 0.01 for the Sina Weibo dataset and 0.001 for the Twitter dataset. The model was trained after 200 epochs for the Sina Weibo and 500 epochs for the Twitter dataset.
We implemented our model by using pytorch\footnote{Our code is available at: https://bit.ly/2UMPtsi}

\begin{figure*}[]
	\begin{subfigure}{0.45\textwidth}
		\centering\includegraphics[width=0.9\textwidth]{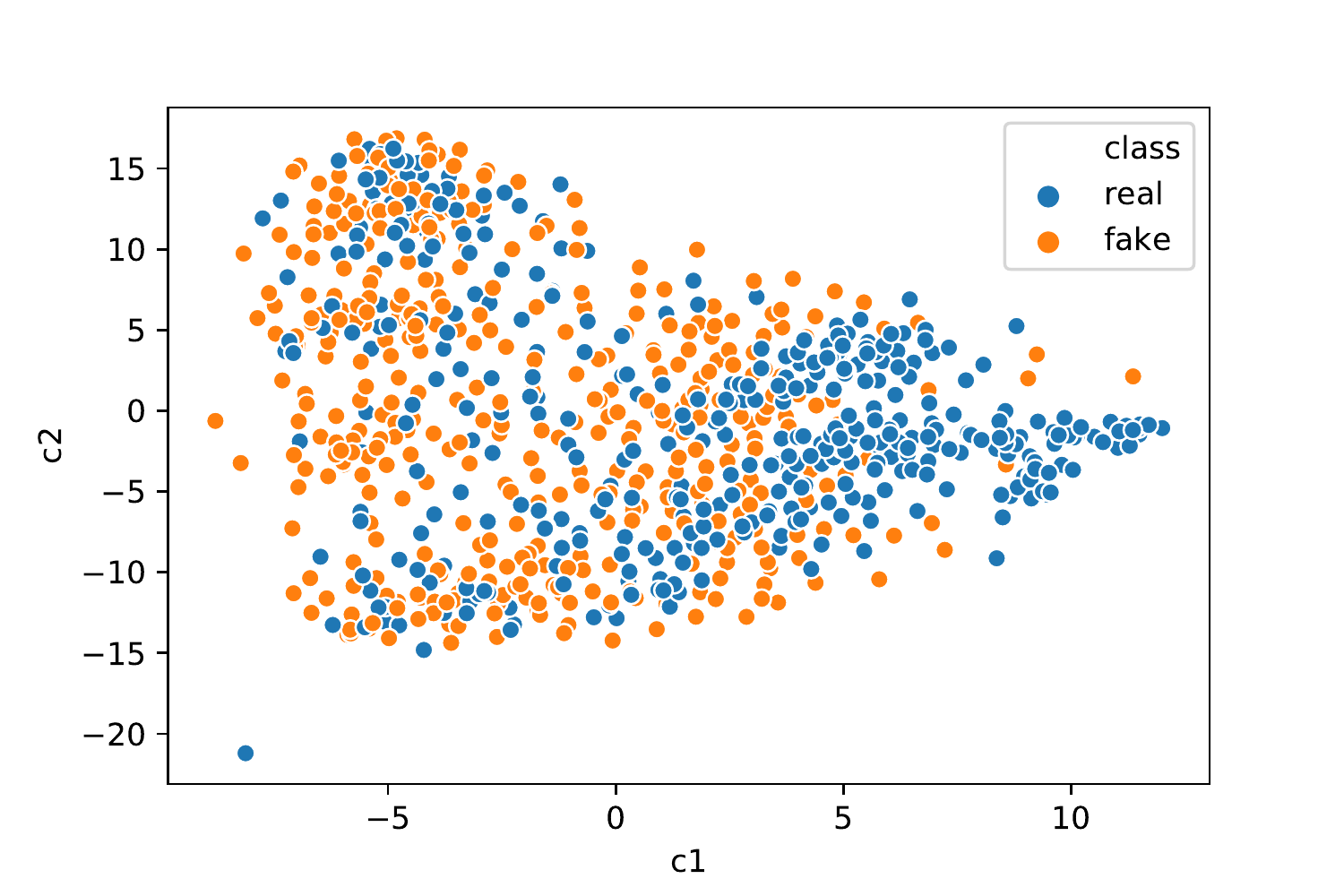}
		\caption{Twitter Input of Model}
		\label{fig:pca_input}
	\end{subfigure}
	\begin{subfigure}{0.45\textwidth}
		\centering\includegraphics[width=0.9\textwidth]{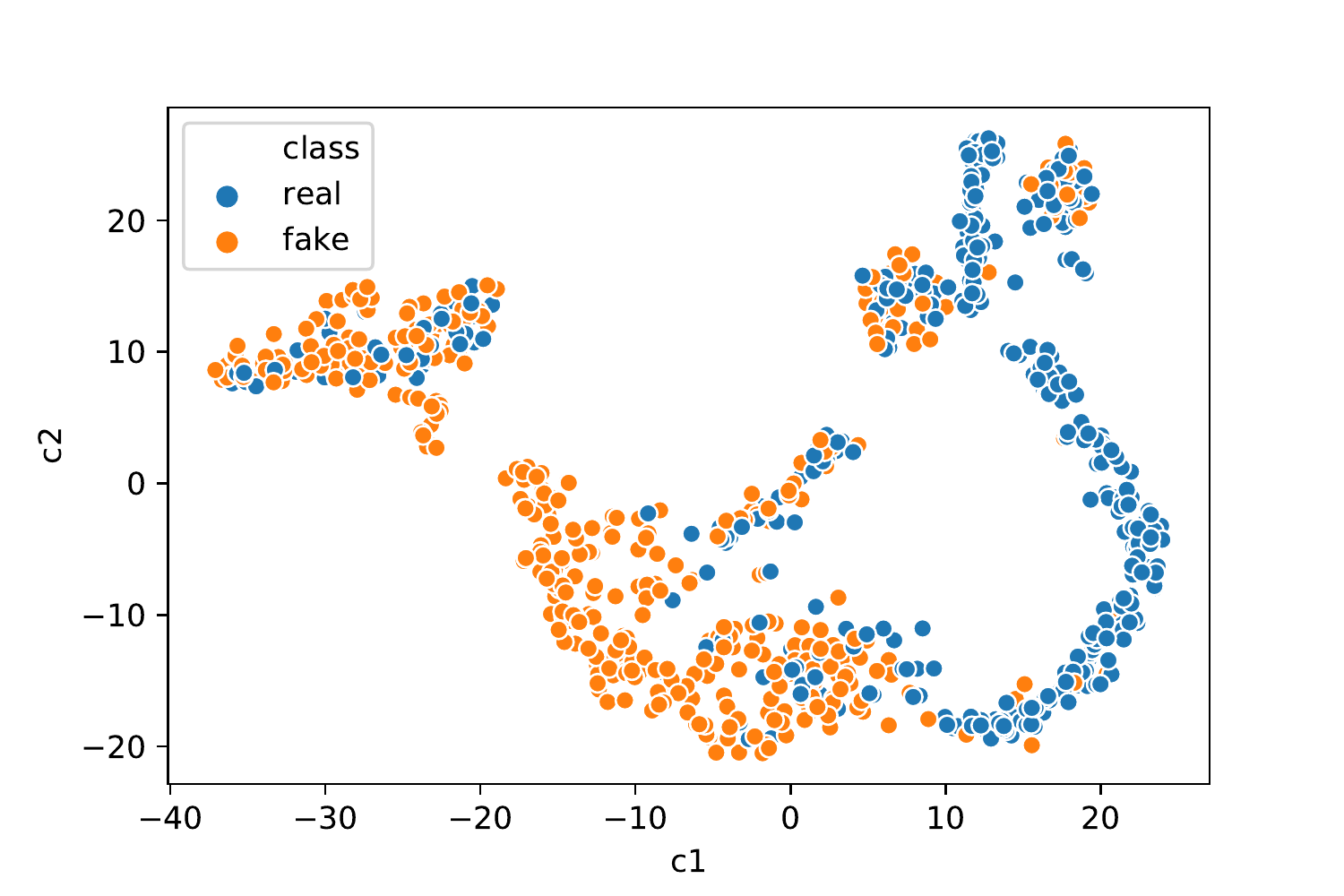}
		\caption{ Twitter Output of Model}
		\label{fig:pca_ouput}
	\end{subfigure}
	\caption{Visualizing Transformations Effects on Input data in Twitter Dataset}
	\label{fig:pca}
\end{figure*}

\subsection{Experimental Results}
\begin{figure}[!]
	\center
	\includegraphics[width=7cm]{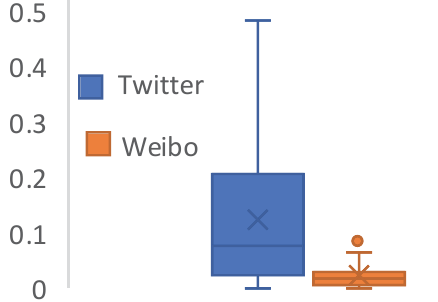}
	\caption{NEC earliness on Twitter and Sina Weibo datasets }
	\label{fig:early}
\end{figure}

The correctness of labeling results is shown in Table \ref{tab:results}. For a fair comparison between models, we set the input length of baseline models similar to our model stopping point. As illustrated, the NEC model outperforms previous models in terms of  news labeling correctness. In Figure \ref{fig:acc}, the accuracy against earliness of NEC is compared to other models. As shown, NEC stops the labeling at about 5 time steps for Sina Weibo and 20 time steps for Twitter datasets, while other models use the whole message stream to achieve their labeling. Despite this significant advantage, NEC has a close competition with other models in term of accuracy. The earliness metric distribution for different events is shown in Figure \ref{fig:early}. As depicted, despite the existence of variable event lengths, the earliness variance is low.

\subsection{Model Analysis}
As shown in Figure \ref{fig:lossepoch}, the loss is decreasing as the model training proceeds in epochs. The proposed model objective function is a combination of accuracy and earliness. It is seen that earliness is improving as training progresses. At the beginning of learning phase a complete sequence is needed for model to label the news, and the earliness is equal to 1. However, after some training, earliness enhances and reaches to about 0.19 for Twitter and 0.05 for Sina Weibo. After this stage, earliness will not improve, because it reaches to a point that decreasing the detection length will have a negative impact on accuracy.
\begin{figure}
	\center
	\includegraphics[width=0.5\textwidth]{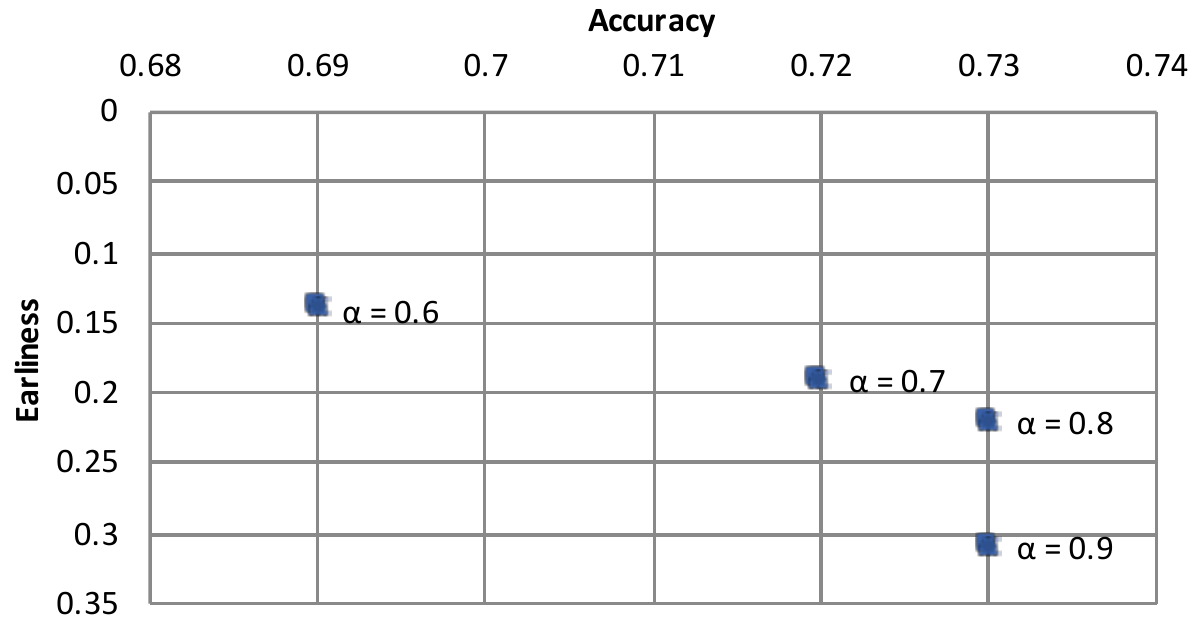}
	\caption{Effect of hyper parameter $\alpha$ on earliness and accuracy.}
	\label{fig:alpha}
\end{figure}
The effect of hyperparameter $\alpha$ in the stopping rule (Equation \ref{eq:stop}) on accuracy and earliness is shown in Figure \ref{fig:alpha}. Clearly, $\alpha$ should be greater than 0.5 meaningful selection of labels. When $\alpha$ is 0.6, the accuracy decreases because the probability of 0.6 is not sufficient for labeling, however earliness improves as expected. By increasing the value of $\alpha$ to 0.7 or 0.8, the accuracy increases. However, after that point, i.e. for $\alpha$ equal to 0.9, the accuracy would not improve any more. Thus we can claim that after reaching the probability of 0.8, the labeling of learned model is not crediable.
To gain a better insight on how the model works, we used PCA and t-sne \cite{vanDerMaaten2008} to decrease the dimension of input data and output of model, for the Twitter dataset. As shown in Figure \ref{fig:pca_input}, the input data is not separable. However, as shown in Figure \ref{fig:pca_ouput}, after model transformation on the input data, it can be easily separated.

\section{Conclusions and Future Work}

In this paper, we introduced a new real time early news labeling method called NEC. We proposed a novel loss function for training a recurrent neural network, and a new stopping rule for finding earliest consistent labeling point. Moreover, the news representation used for classification is composed of publicly available information without utilizing network inference or any prior knowledge. Experiments on real datasets demonstrates that NEC outperforms the competitive methods in term of accuracy while detecting in an earlier stage. As the future works, we may evaluate the performance of adding an attention mechanism to the current model.
{Also we are interested in combining RNN with graphical models for capturing the local dependencies can be . Moreover, we plan to extend our model to semi-supervised classification. Considering multi-modal data (both the text and image information simultaneously) for early fake news detection is an interesting direction for this work.}

\pagebreak

\bibliographystyle{IEEEtran}
\bibliography{IEEEabrv,ref}
\end{document}